\documentclass[twoside]{article}
\usepackage{fleqn,espcrc2,epsfig}
\title{$\varepsilon '/\varepsilon$ in the Standard Model}
\author{I. Scimemi
\address{Departament de F\'{\i}sica Te\`orica, IFIC, 
  Universitat de Val\`encia -- CSIC\\ E-46100 Burjassot (Val\`encia), Spain}
\thanks{Talk presented at QCD2000, Montpellier july 2000.
I  warmly thank  the  collaboration of E. Pallante and A. Pich 
and the  discussions with J. Portoles.
This work has been supported in part by the European Union 
TMR Network  
``EURODAPHNE'' (Contract No. ERBFMX-CT98-0169).
Report: IFIC/00-65.}
}

\newcommand{\eps}{\varepsilon}

\newcommand{\as}{\alpha_{s}}  
\newcommand{\bea}{\begin{eqnarray}}  
\newcommand{\eea}{\end{eqnarray}}  
\newcommand{\beq}{\begin{equation}}  
\newcommand{\eeq}{\end{equation}}  
\newcommand{\nn}{\nonumber}  
\newcommand{\fr}{\frac}

\newcommand{\ra}{\rightarrow}

\newcommand{\pr}{Phys.~Rev. }  
\newcommand{\pl}{Phys.~Lett. }  
\newcommand{\np}{Nucl.~Phys. }  
  
\newcommand{\cO}{{\cal O}} 
\newcommand{\cL}{{\cal L}} 
\newcommand{\cQ}{{\cal Q}} 
 
\newcommand{\lsim}{~{}_{\textstyle\sim}^{\textstyle <}~} 
\newcommand{\ba}{\begin{array}{c}} 
\newcommand{\bat}{\begin{array}{cc}} 
\newcommand{\ea}{\end{array}} 
\def\eqn#1{(\ref{#1})} 
\def\Journal#1#2#3#4{{#1} {\bf #2}, #3 (#4)}

\def\PLB{{\em Phys. Lett.}  B}

\usepackage{amssymb}
\begin{document}
\begin{abstract}
In order  to provide an estimate of $\eps '/\eps$ several effective theories 
and physical effects have to  be disentangled.
In this talk I discuss how  it is possible to predict $\eps '/\eps$ 
taking into account all sources of large logs.
 The  numerical result one obtains, 
$\eps '/\eps \sim (1.7\pm 0.6) \cdot 10^{-4}$,
 is  in good agreement with present measurements.
\end{abstract}

\maketitle
\section{Introduction}

\begin{figure}[tbh]       
\setlength{\unitlength}{0.5mm} \centering 
\begin{picture}(165,120) 
\put(0,0){\makebox(165,120){}} 
\thicklines 
\put(10,105){\makebox(40,15){\large Energy Scale}} 
\put(58,105){\makebox(36,15){\large Fields}} 
\put(110,105){\makebox(40,15){\large Effective Theory}} 
\put(8,108){\line(1,0){149}} {\large 
\put(10,75){\makebox(40,27){$M_W$}} 
\put(58,75){\framebox(36,27){$\ba W, Z, \gamma, g \\ 
     \tau, \mu, e, \nu_i \\ t, b, c,\dots \ea $}} 
\put(110,75){\makebox(40,27){Standard Model}} 
 
\put(10,40){\makebox(40,18){$\lsim m_c$}} 
\put(58,40){\framebox(36,18){$\ba  \gamma, g   ;\, \mu ,  e,    
             \\ \nu_i;\, s, d, u \ea $}}  
\put(110,40){\makebox(40,18){$\cL_{\mathrm{QCD}}^{(n_f=3)}$, \   
             $\cL_{\mathrm{eff}}^{\Delta S=1,2}$}} 
 \put(10,5){\makebox(40,18){$M_K$}} 
\put(58,5){\framebox(36,18){$\ba\gamma  ;\, \mu , e, \nu_i  \\  
            \pi, K,\eta  \ea $}}  
\put(110,5){\makebox(40,18){ChPT}} 
\linethickness{0.3mm} 
\put(76,37){\vector(0,-1){11}} 
\put(76,72){\vector(0,-1){11}} 
\put(80,64.5){OPE}  
\put(80,29.5){$N_C\to\infty$}} 
\end{picture} 
\caption{Evolution from $M_W$ to the kaon mass scale. 
  \label{fig:eff_th}} 
\end{figure}
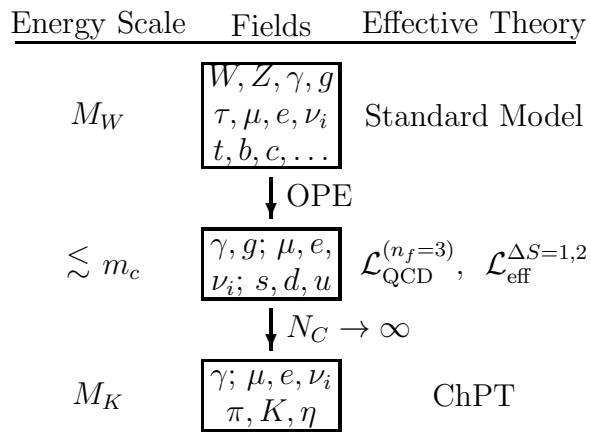

$\eps '/\eps $  is  a fundamental  
test for our understanding of flavor--changing phenomena. 
It represents a great source of inspiration for physics research 
and has motivated in recent years a very interesting scientific 
controversy, both on the experimental and theoretical sides. 
The present world average is~\cite{na48}  
\beq  
{ \rm Re} \left(\eps '/\eps\right) =(1.93 \pm 0.24) \cdot 10^{-3} \ 
,  
\label{eq:ave}
\eeq  
providing clear evidence for a non-zero $\eps '/\eps $ value.
The theoretical status is instead more debated 
(see ref.~\cite{Trieste} for a brief review).
 One  of  the problems that had to be faced during the past  years
  has been  that,
 while   CP violation  is born at a scale say ${\cal O}(M_W)$,
 the observables that enter  in  the game  have to be estimated 
 at  a scale of    ${\cal O}(M_K)$ (see fig.~\ref{fig:eff_th}).
Changing the  order of magnitude of the scales one considers,
  different physical effects
 appear and have to be disentangled.
 In this picture, the physics is described by a chain of different 
effective field theories, with different particle content, which match  
each other at the corresponding boundary (heavy threshold). This 
procedure permits to perform an explicit summation of large logarithms 
$t\equiv\ln{(M/m)}$, where $M$ and $m$ refer to any scale 
appearing in the evolution.

One gets finally an effective $\Delta S=1$ Lagrangian, defined in the 
three--flavor theory \cite{GW:79,BURAS}, 
\beq\label{eq:Leff}  
 {\cal L}_{\mathrm eff}^{\Delta S=1}= - \frac{G_F}{\sqrt{2}} 
 V_{ud}^{\phantom{*}}\,V^*_{us}\,  \sum_i  C_i(\mu) \; Q_i (\mu) \; ,  
 \label{eq:lag}  
\eeq  
which is a sum of local four--fermion operators $Q_i$, 
constructed with the light degrees of freedom, modulated 
by Wilson coefficients $C_i(\mu)$ which are functions of the 
heavy masses.  
We have explicitly factorized the Fermi coupling $G_F$ 
and the Cabibbo--Kobayashi--Maskawa (CKM) matrix elements 
$V_{ij}$ containing the usual Cabibbo suppression of $K$ decays. 
 
The overall renormalization scale $\mu$ separates 
the short-- ($M>\mu$) and long-- ($m<\mu$) distance contributions, 
which are contained in $C_i(\mu)$ and $Q_i$, respectively. 
The physical amplitudes are of course independent of $\mu$; thus, the  
explicit scale/scheme dependence of the Wilson coefficients should 
cancel exactly with the corresponding dependence of the $Q_i$ 
matrix elements between on-shell states.
Usually  one refers to this as the ``matching'' between Wilson coefficients 
and   hadronic matrix elements.
Thanks to the completion of the next-to-leading 
logarithmic order calculation of the Wilson coefficients 
\cite{buras1,ciuc1}, 
all gluonic corrections of $\cO(\as^n t^n)$ and $\cO(\as^{n+1} t^n)$ 
are already known. Moreover, the complete $m_t/M_W$ dependence (at lowest 
order in $\as$) has been taken into account. 
We will fully use this information up to scales $\mu\sim \cO(1\; {\rm GeV})$, 
without making any unnecessary expansion in powers of $1/N_c$. 
 The most debated part of the calculation regards the estimate
 of hadronic matrix elements. In the following it is presented a  
brief  summary of the main ingredients of this part of the computation.

\section{Low energy effective theory}

In order to define an effective field theory at low  scale one can
 use  global symmetry considerations.
 In this way one arrives   at the $\chi$PT description of the Standard Model,
 which describes  the dynamics of the QCD Goldstone bosons ($\pi$, $K$, $\eta$)
  as  an expansion  in powers of momenta and quark masses over the chiral
 symmetry breaking scale ($\Lambda_\chi \sim 1$ GeV).

At lowest order, 
the relevant most general effective bosonic Lagrangian, with the same 
$SU(3)_L\otimes SU(3)_R$ transformation properties and quantum numbers 
as the short--distance Lagrangian \eqn{eq:Leff}, contains four terms: 
\bea \nn
\cL_2^{\Delta S=1} &=& -{G_F \over \sqrt{2}}  V_{ud}^{\phantom{*}} V_{us}^* 
\Bigg\{ g_8\; f^4  \;\langle\lambda L_{\mu} L^{\mu}\rangle \Biggr. \\
\nn 
& & +
g_{27}\; f^4 \,\left( L_{\mu 23} L^\mu_{11} + {2\over 3} L_{\mu 21} L^\mu_{13} 
\right)  
 \nn\\ &&+ e^2 f^6 g_{EW} \;  
\langle\lambda U^\dagger \cQ U\rangle \Bigg\} + 
\mbox{\rm h.c.} \, ,  
\label{eq:lg8_g27}
\eea 
where the matrix $L_{\mu}=-i U^\dagger D_\mu U$  represents the octet of  
$V-A$ currents, at lowest order in derivatives, 
$f\sim f_\pi=92.4 $ MeV,
$\cQ= {\rm diag}(\frac{2}{3},-\frac{1}{3},-\frac{1}{3})$ is the quark 
charge matrix,
 $\lambda\equiv (\lambda^6 - i \lambda^7)/2$ projects onto the 
$\bar s\to \bar d$ transition [$\lambda_{ij} = \delta_{i3}\delta_{j2}$]
and $\langle {\mbox{A}}\rangle$ denotes the flavor trace of A.
The chiral couplings $g_8$ and $g_{27}$ measure the strength of
 the two 
parts of the effective Hamiltonian \eqn{eq:Leff} transforming as  
$(8_L,1_R)$ and $(27_L,1_R)$, respectively, under chiral rotations. 
The moduli   of $g_8$ and $g_{27}$ can be extracted from 
 the CP--conserving part of 
$K \rightarrow 2 \pi$ decays; at lowest order a 
phenomenological analysis gives 
\cite{PGR:86}: 
%
$\left| g_8 +{1\over 9}\, g_{27}\right| \simeq 5.1$,
$\left| g_{27}/ g_8 \right| \simeq 1/18$. 
%
The huge difference between these two couplings 
shows the well--known 
enhancement of the octet $\vert\Delta I\vert = 1/2$ transitions.

 The theoretical  calculation of the couplings $g_I$  is a difficult task.  
 One observes, however, that  what really matters  in the calculation of  
$\eps '/\eps$   are  not 
 the moduli  of these couplings $g_I$ but 
 their imaginary parts.
 In fact as ${\rm Im} g_I\ll{\rm Re} g_I $, one  can deduce ${\rm Re} g_I$
  from  the experiment, but 
 it is necessary to give  a prediction for ${\rm Im} g_I$.
 Moreover the matching between the  Lagrangians of eq.~\ref{eq:Leff} 
and eq.~\ref{eq:lg8_g27} must be provided.
The large--$N_c$ expansion offers the possibility  to solve both  these
 problems in a simple  and elegant way.
In the limit of a large color number  each four--quark operator
 factorize into currents which have  a well known chiral realization.
 Thus one obtains
\bea
g_8^\infty&=& -{2\over 5}\,C_1+{3\over 5}\,C_2+C_4-{16B_0^2\over f^2}\,L_5\,C_6
\ ,\nonumber\\
g_{27}^\infty&=&{3\over 5}\,(C_1+C_2)\ ,\nonumber\\
g_{EW}^\infty&=& -{3B_0^2\over e^2f^2}\, C_8\ ,
\label{eq:c2}
\eea
 where $B_0=-\langle \bar{q} q\rangle (\mu)/f_\pi^2$.
 Now the dominant part of the contributions to $\eps '/\eps$
(or, which is the same, to ${\rm Im}g_{8,EW}$)  is provided 
by the operators $Q_{6,8}$ whose behavior in the large--$N_c$ limit  is
different from the rest of operators.
 In fact when $N_c\rightarrow \infty$ the only anomalous dimensions 
which survive are  the ones corresponding  to these 
operators~\cite{BG:87,BBG:87}. 
Then $Q_{6,8}$ factorize  in the product of color--singlet scalar and
pseudoscalar currents  which  generate the factors $B_0$ of eq.~\ref{eq:c2}.
 The 
 scale/scheme dependence of the quark condensate $B_0$ exactly cancels 
the one of the Wilson coefficient $C_{6,8}$~\cite{BBG:87}.
 The anomalous dimension of all the other operators is zero for 
$N_c\ra \infty$.  This means  that in order to achieve  a reliable estimate
 of the matrix elements  of these operators it is necessary to go to 
next--to--leading order  in the $N_c$ expansion.
That is why the  $\Delta I=1/2$ rule  is so difficult to demonstrate.
Finally one  notes that eq.~\ref{eq:c2} is perfectly equivalent to say 
$B_8^{(3/2)}\sim B_6^{(1/2)}=1$. That is, upto minor  inputs, the
prediction obtained in both large--$N_c$  and $\chi$PT expansions reproduces 
the results of ref.~\cite{buras1,ciuc1}.
\section{Chiral loops}
The ultraviolet logarithms that have been resummed using the renormalization 
group equation are not the only source  of large logs appearing 
in the estimate of   $\eps '/\eps$.
 It is well known~\cite{KA91} that infrared logs    provide 
 a   source of 
enhancement for $I=0$ amplitudes  which has to be taken into account.
The one loop correction already provides an enhancement of
 about $40\%$  and still underestimates
  the observed  $\delta_0^0$ phase shift.
A resummation of higher order effect is so necessary and it has been provided 
in ref.~\cite{PP:00,PP:00b} and discussed  also in this 
conference~\cite{pall:qcd}.
 The  approach is based on  the Omn\`es solution for $K\ra\pi\pi$
 amplitudes~\cite{ALLOMNES}.
The Omn\`es solution for a CP conserving (but the same conclusions hold
 also for the CP violating)  $K\to\pi\pi$ amplitude can be 
written in the generic form 
\bea 
&&\hspace{-0.6cm}{\cal A}_I \, =   
\left(M_K^2-M_\pi^2\right) \; a_I(M_K^2) \nonumber\\
&&\, =\,   
\left(M_K^2-M_\pi^2\right) \; \Omega_I(M_K^2,s_0) \; a_I(s_0)\nn \\
&&\, = \,
\left(M_K^2-M_\pi^2\right) \; \Re_I(M_K^2,s_0) \; a_I(s_0) 
\; e^{i\delta^I_0(M_K^2)}\, , 
\nn
\eea 
where $a_I(s)$ as a function of the total energy squared $s$ has been
 explicitly computed up to one--loop in ChPT \cite{PP:00b}.
The Omn\`es factor $\Omega_I(M_K^2,s_0)$ can be interpreted as an 
evolution operator from the subtraction point $s_0$ to $M_K^2$. 
It can be split into the dispersive 
contribution $\Re_I(M_K^2,s_0) $ and the usual phase shift exponential. 
For each of the amplitudes $a_I$, with $I=0,2$ 
the $s$ dependence can be written in a simple form:
\beq\label{eq:s_dep} 
a_I(s) = a_I(0)\;\left\{ 1 + g_I(s) 
  + O(p^4)\right\}\, . 
\eeq
 The $s$ dependence of the one--loop correction at low values of $s$ 
is dominated by the pure $SU(2)$ effect of elastic $\pi\pi\to\pi\pi$ 
scattering. These universal infrared effects enhance the $I=0$ amplitudes 
while suppress the $I=2$ amplitude. 
I underline that  the only
role of the Omn\`es factor remains that of providing an
efficient resummation of large infrared effects due to FSI. The advantage
of the Omn\`es exponentiation respect to the usual one--loop ChPT 
computation is to control the uncertainty coming from higher order
($\geq$ two--loops) FSI effects.

Taking a low subtraction point $s_0 =0$ where higher--order 
corrections are expected to be small, we can just multiply the tree--level 
 amplitudes with the experimentally determined Omn\`es 
exponentials \cite{PP:00b}.
The two dispersive correction factors thus obtained are
\bea
\Re_0(M_K^2,0)&=& 1.55 \pm 0.10;\nn \\
\Re_2(M_K^2,0) &=& 0.92 \pm 0.03\, ,
\label{DISP}
\eea
where the errors take into account a) the uncertainties of the 
fits to the experimental phase shifts data used in the calculation of the 
Omn\`es factor and b) 
the additional inelastic contributions above the first inelastic threshold.
Finally since FSI effects are next-to-leading in the $1/N_c$ expansion 
but numerically
large, this procedure avoids any double counting.
Since the Omn\`es factor can be applied directly to each matrix 
element $\langle Q_i\rangle_I$, the final estimate with the inclusion of 
FSI effects can be expressed via the product
$
\langle Q_i\rangle_I = \langle Q_i\rangle_I^\infty\, \cdot \Re_I\, .
$
\section{Numerical results}
 A full numerical  analysis  of $\eps'/\eps$ taking into account
also smaller  effects will be  presented elsewhere~\cite{nos}.
 To a reasonably good approximation however one can take as 
an estimate the one coming from~\cite{buras1}
\bea
\fr{\eps '}{\eps} &\sim &\left[B_6^{(1/2)} (1-\Omega_{IB})-0.4 B_8^{(3/2)}\right] \ ,
\eea
 where now the B--parameters and the factor $\Omega_{IB}$
 have to be corrected taking into  account 
also the contribution of FSI, that is 
\bea
B_6^{(1/2)} &=&B_6^{(1/2)}|_{N_c\ra\infty}\cdot \Re_0(M_K^2,0)= 1.55 \ ,\nn\\
B_8^{(3/2)} &=&B_8^{(3/2)}|_{N_c\ra\infty}\cdot \Re_2(M_K^2,0)= 0.92 \ ,\nn\\
\Omega_{IB} &=&0.16 \cdot\Re_2(M_K^2,0)/\Re_0(M_K^2,0)=  0.09 \ .\nn
\eea
The  effect of FSI is  numerically evident.
 The cancellation between the  $I=0$ and $I=2$ amplitudes is  strongly removed
 and also the effect of the isospin breaking term proportional to $\Omega_{IB}$
 looses its weight.
 The prediction one obtains  for $\eps '/\eps$ is so
\beq
\eps'/\eps \simeq (1.7 \pm 0.6) \cdot 10^{-4}
\eeq
which  compares well with the present world average in eq.~\ref{eq:ave}.
In this prediction all sources of large logs
 are finally taken into account. 
$Q_{6,8}$ are well approximated by  the leading terms in $1/N_c$.
Therefore one expects reasonably that the size of the missing NLO--$N_c$ 
corrections are of the order of $30\%$. 
At present the estimate of  these effects can only be done within specific 
models~\cite{Trieste,Dortmund,BP:00,PR:91,K:99}.

\end{document}